\documentclass[pdflatex,sn-mathphys-num,iicol]{sn-jnl}

\usepackage{graphicx}%
\usepackage{multirow}%
\usepackage{amsmath,amssymb,amsfonts}%
\usepackage{amsthm}%
\usepackage{mathrsfs}%
\usepackage[title]{appendix}%
\usepackage{xcolor}%
\usepackage{textcomp}%
\usepackage{manyfoot}%
\usepackage{booktabs}%
\usepackage{algorithm}%
\usepackage{algorithmicx}%
\usepackage{algpseudocode}%
\usepackage{listings}%
\usepackage{braket}
\usepackage{tabularx}
\usepackage[numbers,sort&compress]{natbib}   

\begin{document}

\title{Cryogenic Systems for Quantum Photonic Technologies: A Practical Review}


\author[1,2]{\fnm{Alex H.} \sur{Rubin}}\email{ahrubin@ucdavis.edu}

\author[1,2,3]{\fnm{Victoria A.} \sur{Norman}}

\author[2]{\fnm{Marina} \sur{Radulaski}}

\newcommand{\killperiod}[1]{}

\affil[1]{\orgdiv{Department of Physics}, \orgname{University of California, Davis}, \orgaddress{\street{1 Shields Avenue}, \city{Davis}, \postcode{95616}, \state{CA}, \country{USA\killperiod}}}

\affil[2]{\orgdiv{Department of Electrical and Computer Engineering}, \orgname{University of California, Davis}, \orgaddress{\street{1 Shields Avenue}, \city{Davis}, \postcode{95616}, \state{CA}, \country{USA\killperiod}}}

\affil[3]{\orgdiv{Cavendish Laboratory}, \orgname{University of Cambridge}, \orgaddress{\street{J.J. Thomson Avenue}, \city{Cambridge} \postcode{CB3 0HE}, \country{United Kingdom}\killperiod}}


\abstract{

While nonclassical light sources are fundamental to quantum communication and computing, solid-state platforms like color centers and quantum dots require cryogenic temperatures to reach the performance levels necessary for practical applications.
Over the past decade, low-temperature engineering has transitioned from manual handling of liquid cryogens to automated closed-cycle cryostats.
This review details the principles behind modern cooling hardware ranging from flow cryostats to mechanical cryocoolers and dilution refrigerators, with a specific focus on the requirements of optical quantum devices.
Aimed at the practicing scientist, this overview provides the technical insights and historical context needed to navigate the current cryogenic landscape and evaluate its role in the future of quantum technology deployment.
}


\maketitle

\section{The importance of cryogenics for optical quantum technology}
The past two decades have seen huge leaps in quantum information technologies, which harness entanglement and superposition to outperform classical limits in computation, communication, and sensing.
These advantages can only be realized if the quantum states involved remain coherent and undisturbed.
In most platforms, thermal fluctuations in the environment constitute a primary source of decoherence, driving both energy relaxation and pure‑dephasing processes that ultimately limit coherence times and device performance.
To mitigate this decoherence, most quantum systems critically depend upon cryogenic refrigeration to cool them to extremely low temperatures, where thermal excitations are minimal, and the long coherence times needed for quantum operations can be achieved.

Virtually any system which exhibits quantum behavior can in principle be used to store and process quantum information.
As a result, a huge number of hardware platforms have been explored for this purpose, including electronic or nuclear spins, superconducting Josephson junctions, mechanical oscillators, and free electrons on liquid helium, to give just a few examples.
However when it comes to communicating quantum information over long distances, photons are nearly unchallenged.
The availability of ultra low-loss optical fiber and the fact that photons do not require cryogenics to maintain coherence combine to make them the ideal ``flying qubit" for use in quantum networks and quantum computers.
However, while photons themselves are robust to decoherence, the matter-based emitters used to generate them (such as quantum dots or color centers) are inherently exposed to decohering interactions with their solid-state host, necessitating low temperatures to serve as bright sources of identical single photons.

The optical transitions of these emitters are generally coupled to phonons in their environment.
The involvement of phonons in the transition causes the emitted photon to vary, resulting in non-identical photons.
The optical spectrum of these solid-state emitters can be broken into two parts: the zero-phonon line (ZPL), resulting from optical transitions in which no phonons were involved, and the phonon sideband (PSB), produced when a phonon did affect the transition.
For quantum technologies the ZPL is the critical feature, as photons emitted through this channel are spectrally pure and coherent, enabling the indistinguishability required for quantum interference and entanglement.
The brightness of the ZPL as a fraction of the total emission is quantified by the Debye–Waller factor, which is a key metric for assessing an emitter’s usefulness.
Because phonon interactions broaden the emission and reduce coherence, most solid-state quantum emitters must be operated at cryogenic temperatures.
Cooling suppresses phonon populations, thereby narrowing the ZPL linewidth and increasing the fraction of photons emitted into the ZPL.

Some emitters also possess an internal spin degree of freedom that can serve as a stationary qubit.
Spin-dependent fluorescence enables conversion between the spin state and a photonic state, or generation of spin–photon entanglement.
Such spin–photon interfaces already underpin commercial quantum sensors and hold broad potential for scalable quantum networks and nonlocal connectivity in quantum processors.
Like all matter-based qubits, these spin systems are susceptible to decoherence from interactions with their surrounding lattice and electromagnetic environment.
Maintaining long spin coherence times therefore requires cryogenic operation, which suppresses phonon-driven relaxation and spin dephasing processes.

When it comes to detecting single photons, the state of the art is represented by superconducting nanowire single-photon detectors (SNSPDs).
These devices offer near-unity detection efficiency, low timing jitter, and dark-count rates orders of magnitude below those of semiconductor detectors, making them indispensable for quantum communication and photonic quantum computing.
SNSPDs operate by biasing an ultrathin superconducting wire just below its critical current; the absorption of a single photon locally drives the wire into a resistive state, producing a measurable voltage pulse.
To maintain superconductivity and minimize thermal noise, the detector must be cooled well below its critical temperature (typically below 3 K) using cryogenic refrigeration.
As a result, even on the detection side, cryogenics remain essential to realizing high-performance optical quantum technologies.

Cryogenic cooling systems, like conventional refrigerators, rely on the phase change of a working fluid.
Helium, the lightest noble gas, has the lowest boiling point of any substance (4.2 K) and is therefore central to cryogenics.
Early cryostats used liquid helium baths to reach 4.2 K; temperatures down to about 1 K could be achieved by using a vacuum pump to lower the vapor pressure above the liquid.
To access the sub-kelvin regime, the rare isotope helium-3 is required.
Mixtures of the two isotopes exhibit a unique phase transition at low temperatures, which enables dilution refrigerators in which helium-3 dissolves into helium-4, allowing access to temperatures in the milliKelvin range.
Modern cryostats used in optical quantum experiments are specifically engineered to provide ultralow temperatures with excellent optical access, minimal vibration, and compatibility with strong magnetic fields required by some emitters.
Designs are optimized to support free-space or fiber coupling and to integrate electrical feedthroughs for device control.

This tutorial introduces the main cryogenic systems used in contemporary quantum optics experiments and explains their design considerations, operating principles, and performance characteristics.



\begin{figure}[htp]
    \centering
    \includegraphics[width=\columnwidth]{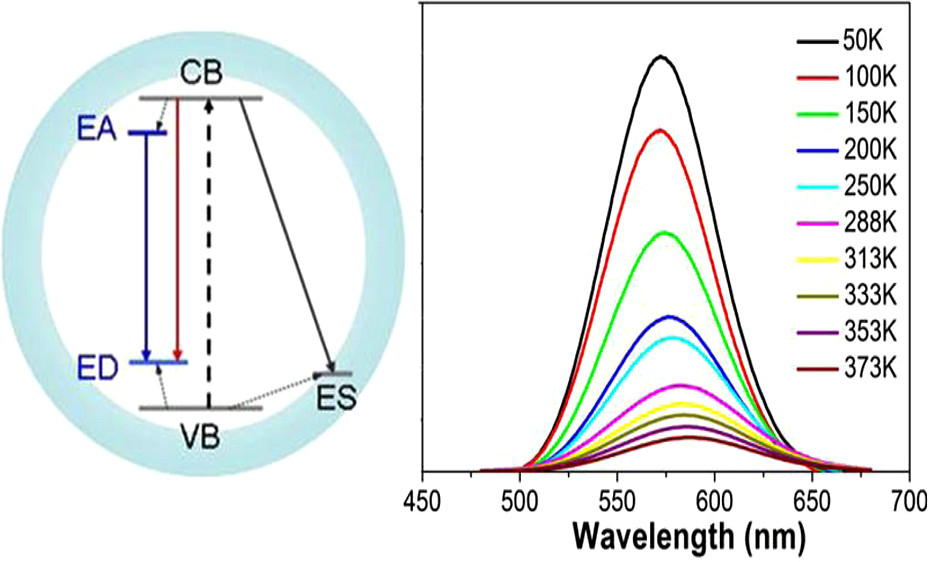}
    \caption{
        Photoluminescence spectra of a ZnCuInS/ZnSe/ZnS quantum dot at temperatures ranging from 50 K to 373 K \cite{Liu2013}.
    }
    \label{fig:qd-pl}
\end{figure}

\subsection{Quantum dots}
Quantum dots (QDs) are nanoscale semiconductor structures in which a lower-bandgap material is surrounded by higher-bandgap barriers, confining charge carriers in all three spatial dimensions and creating discrete energy levels.
When the confined states support radiative transitions, the QD becomes optically active, enabling high-quality single photon emission.



Because they comprise hundreds to thousands of atoms, QDs have much larger transition dipole moments than atomic-like color centers, making them extremely bright quantum emitters that couple strongly to optical cavities and waveguides.

In contrast to color centers, which have a Debye-Waller factor of a few percent, QDs commonly emit 90\% of their light into the zero-phonon line, reflecting their lower electron-phonon coupling \cite{Denning:20}.
Figure~\ref{fig:qd-pl} shows the temperature-dependent photoluminescence of a ZnCuInS/ZnSe/ZnS quantum dot, illustrating how the emission spectrum broadens and shifts as temperature increases from 50 K to 373 K.

Still, phonons represent a significant dephasing mechanism for the QD spin \cite{PhysRevLett.93.016601}.
Quantum dots can confine single electrons, whose two degenerate spin states may then be split by applying a strong, static magnetic field.
The spin state can then be manipulated by magnetic resonance with radiofrequency pulses, or by ultrafast optical pulses \cite{Press2008}.
Consequently, the system must be cooled to cryogenic temperatures (typically $\sim 4$ K or below) to ensure that the thermal energy is negligible compared to the Zeeman splitting.

\subsection{Color centers}
A color center is a point defect in the crystal lattice of a semiconductor that gives rise to discrete, localized electronic states within the band gap of the host material.
These localized states support sharp optical transitions and, in many cases, host electronic spin degrees of freedom that can be coherently controlled.
Thus a color center acts effectively like a solid-state artificial atom immobilized in bulk crystal.
In close analogy to atomic systems, a single defect emits one photon at a time with high spectral purity and indistinguishability, making color centers an attractive building block for quantum communication and networking.
The associated electronic spin in some color centers exhibits coherence times extending to seconds, enabling long-lived quantum memory with high-fidelity coherent control \cite{5sec_sic}.
In color centers with spin-selective optical transitions, the spin state can be entangled with emitted photons, providing a direct interface between stationary and flying qubits \cite{chu2015quantum}.
Spin-photon entanglement in color centers has recently been used to demonstrate major milestones towards quantum networking \cite{lukin_quantum_network, lukin_quantum_network_nuclear_memory}, blind quantum computing \cite{lukin_blind_qc}, and loophole-free tests of the Bell inequality \cite{Hensen2015}.


\begin{figure}[htp]
    \centering
    \includegraphics[width=\columnwidth]{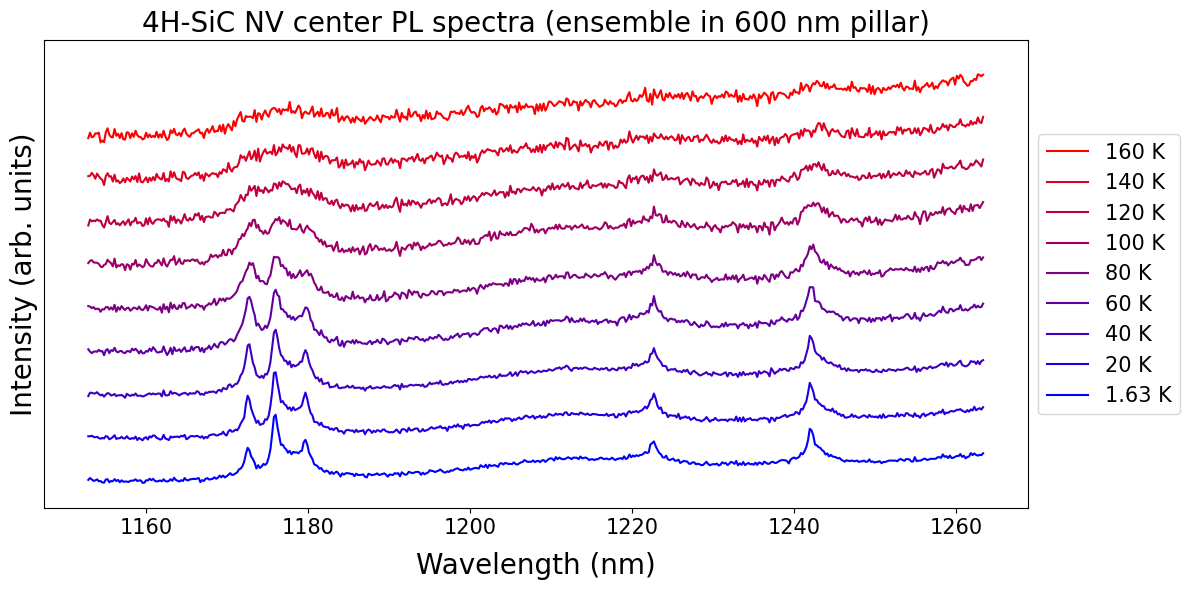}
    \caption{
        The photoluminescence spectrum of an ensemble of NV centers in 4H silicon carbide, as a function temperature.
        At deep cryogenic temperatures, the ZPLs of the NV centers are sharply defined.
        As the temperature rises, the greater occupation of phonon modes shifts most of the NV emission into the phonon sideband, washing out the ZPLs.
    }
    \label{fig:nv_sic_pl}
\end{figure}

Color centers of a given type share the same local atomic configuration and lattice environment, making them nominally identical quantum emitters.
In the absence of strain, photons emitted from two such centers are therefore expected to be indistinguishable in frequency and temporal profile, as required for quantum interference and entanglement-based protocols.
In practice, however, lattice vibrations distort the defect geometry and shift the electronic potential surfaces, producing vibronic coupling between electronic and phonon states that broadens the optical transition and reduces photon indistinguishability.
The resulting emission spectrum comprises a narrow zero-phonon line (ZPL), corresponding to purely electronic transitions, and a broader phonon sideband (PSB) arising from phonon-assisted processes.
Only photons emitted through the ZPL are mutually indistinguishable.
The fraction of emission into this channel increases sharply with decreasing temperature as phonon occupation is suppressed.
Cryogenic operation is therefore essential to achieve narrow linewidths and high photon indistinguishability in most color-center systems.
Cooling not only narrows the optical linewidth but also suppresses spin-lattice relaxation and dephasing driven by phonons, leading to longer spin coherence times and more stable qubit performance.
Figure~\ref{fig:nv_sic_pl} gives an example of this temperature dependence for an ensemble of NV centers in silicon carbide, showing the dramatic narrowing of the zero-phonon line and suppression of the phonon sideband as temperature decreases from 160 K to 1.5 K.

Standard deep-level defects (those whose electronic states are strongly localized to the defect site and lie deep within the band gap) exemplify color centers that perform well under conventional cryogenic conditions.
The diamond NV$^{-}$ center, consisting of a substitutional nitrogen adjacent to a vacant lattice site, hosts a spin-triplet ground state that can be manipulated with microwave fields at its characteristic 2.9 GHz zero-field splitting.
The primary source of spin decoherence in diamond NV$^{-}$ centers is the magnetic noise from nearby $^{13}$C nuclear spins in the lattice; isotopic purification to enhance the $^{12}$C concentration dramatically suppresses this bath, enabling electron spin coherence times reaching 1.8 ms at liquid helium temperatures (around 4 K) \cite{Balasubramanian2009}.
At these temperatures, the 637 nm zero-phonon line provides a bright source of indistinguishable photons for quantum networking applications.
Similarly impressive performance is found in several defects hosted by silicon carbide \cite{rlab_sic_review}.
For example, the divacancy center (adjacent vacant silicon and carbon sites) in silicon carbide has demonstrated spin coherence times as long as 5 seconds under comparable cooling conditions \cite{5sec_sic}.
These examples establish that for many color center platforms, cooling to liquid helium temperatures is sufficient to suppress phonon-induced decoherence and achieve the narrow optical linewidths and long-lived quantum memories required for practical applications.

However, not all color centers are so accommodating.
The so-called Group IV defects in diamond (SiV, GeV, SnV, PbV), in which the exogenous atom sits between two vacant lattice sites, exhibit significantly greater sensitivity to phonons due to orbital degeneracy in their electronic structure.
The $D_{3d}$ symmetry of these defects gives rise to doubly degenerate orbital states that are split only weakly by spin-orbit coupling.
This degeneracy creates an efficient pathway for phonon-mediated relaxation between the orbital branches, leading to rapid decoherence at elevated temperatures.
Cooling to liquid helium temperatures proves insufficient to suppress these processes, and state-of-the-art experiments with SiV centers have required millikelvin temperatures inside dilution refrigerators, where coherence times have reached tens of milliseconds \cite{PhysRevLett.120.053603, PhysRevLett.119.223602, PhysRevLett.132.026901}.
Notably, the same sensitivity to strain that makes Group IV defects susceptible to phonons can also be exploited to lift the degeneracy and inhibit phonon-mediated relaxation.
By engineering static strain fields through external stress, it is possible to extend the spin coherence time of the SiV$^{-}$ center at 4 K to roughly what it would be at 100 mK \cite{Sohn2018}.

\begin{figure}[htp]
    \centering
    \includegraphics[width=\columnwidth]{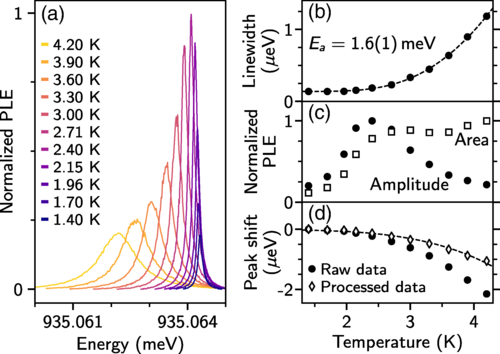}
    \caption{
        ZPL spectral properties of T centers in silicon as a function of temperature.
        As a weakly bound exciton, the T center's optical emission is significantly more sensitive to temperature than that of other kinds of quantum emitter.
        Reproduced from \cite{PRXQuantum.1.020301}.
    }
    \label{fig:t_center_pl}
\end{figure}

By contrast with deep-level defects such as the NV center, whose zero-phonon line arises from a transition between two localized defect orbitals, other defects such as the T, W, and G centers in silicon owe their luminescence to bound excitons \cite{Komza2024, PhysRevB.98.195201}.
The observed zero-phonon line therefore corresponds to excitonic recombination, not a single-particle transition between localized orbitals.
Because the binding energy of these defect-bound excitons is typically only a few tens of meV, thermal dissociation is rapid at room temperature, and no sharp zero-phonon line is observed at all at temperatures above a few Kelvin (see \ref{fig:t_center_pl}) \cite{first_princip_t_center}.
This represents an even more stringent temperature requirement than the Group IV defects, as bound exciton centers simply do not function as quantum emitters without cryogenic cooling \cite{PRXQuantum.1.020301}.
Nevertheless, T centers offer long electron and nuclear spin coherence times and convenient optical emission in the telecom O-band, making them very promising for quantum information applications that can accommodate the necessary cryogenic infrastructure \cite{SimmonsPhotonicArchitecture}.

Beyond temperature control, many color center experiments also require the application of static magnetic fields to the sample.
For a spin-bearing color center, adding a static magnetic field introduces the Zeeman interaction which controllably shifts and splits the spin sublevels.
The direction of the B field sets a quantization axis, and the degree of the splitting can render each allowed spin transition spectrally distinct for selective microwave or radiofrequency driving.
For defects with degenerate spin states at zero field, such as the T center in silicon, an applied magnetic field is essential to break the degeneracy and enable spin-selective operations.
The magnetic field also plays a critical role in separating contributions from different defect orientations: crystallographically equivalent defects with different orientations in the lattice respond differently to an applied field, allowing individual addressability \cite{mem_trans_t_center}.
When a magnetic field is applied to a T center, for example, it both induces a Zeeman shift that splits the zero-phonon line into two branches and changes the polarization of the optical dipole, effects that can be exploited to selectively couple or decouple specific defect orientations from optical cavities \cite{Islam2024}.
Moreover, nuclear spins associated with the defect or nearby lattice sites can serve as long-lived quantum memories, and experiments implementing gates between electronic and nuclear spins rely on a global static magnetic field to provide a shared quantization axis \cite{Taminiau2014}.
The field strength and orientation requirements vary significantly between platforms, ranging from tens of millitesla for some applications to several Tesla for others.
These magnetic field requirements present a constraint on cryostat design choices, as illustrated by the superconducting magnet visible in Fig.~\ref{fig:optical-dil-fridge}.

\subsection{2D materials}
Two-dimensional (2D) materials are crystalline solids consisting of a single layer of atoms or a few atomic layers bound by van der Waals forces.
The reduced dimensionality of these materials dramatically alters their electronic and optical properties compared to their bulk counterparts, leading to phenomena such as direct bandgaps, strong exciton binding energies \cite{Liu2019}.
Among the diverse family of 2D materials explored to date, hexagonal boron nitride (hBN) and transition metal dichalcogenides (TMDs) such as WSe$_2$, WS$_2$, MoSe$_2$, and MoS$_2$ have emerged as particularly promising candidates for quantum photonics due to their ability to host localized quantum emitters \cite{Turunen2022}.
These emitters have been less explored relative to quantum dots and color centers, as they have only been shown to exhibit single-photon emission within the past decade, but they are highly promising for quantum photonic applications.

By far the best-studied 2D material is hBN, and its single photon emitters have been shown to possess high brightness and purity, with narrow spectra that are identical across emitters \cite{Tran2016, hbn_identical_emission}.
The main hBN defect of interest to date is the negatively charged boron vacancy (V$_\text{B}^{-}$) which, similarly to color centers in bulk semiconductors, has discrete energy levels that lie within the band gap of hBN.
Like the NV$^{-}$ centers in diamond and silicon carbide, the V$_\text{B}^{-}$ has overall spin-1, which can be optically initialized and coherently controlled via microwaves \cite{hbn_spin_control}.

The existence of single-photon emitters in hBN with spin coherence at room temperature \cite{Stern2024, zhou2026optically, zheng2026surface} could indicate a shift away from needing cryogenics in the future of quantum optics. However, there is much to understand about quantum emitters in hBN including information sometimes as fundamental as the structures of some of these defects themselves. Careful analysis of low temperature spectra provide valuable insight into the physics governing the spectral and spin dynamics of such systems leading to more controllable, scalable quantum information devices.


There are several approaches to integrating these defects into nanophotonic structures.
One simple methods involves placing a flake of hBN on top of photonics fabricated from another material; the atomic thinness of hBN allows defects within it to couple to the evanescent field of the photonic devices below \cite{hbn_on_top, hbn_taper_fiber}.
It is also possible to invert this by fabricating photonic devices on top of an hBN layer \cite{hbn_under_waveguide}.
Perhaps most interestingly, bulk hBN can be directly fabricated into waveguides \cite{hbn_waveguide}.

\begin{figure}[htp]
    \centering
    \includegraphics[width=\columnwidth]{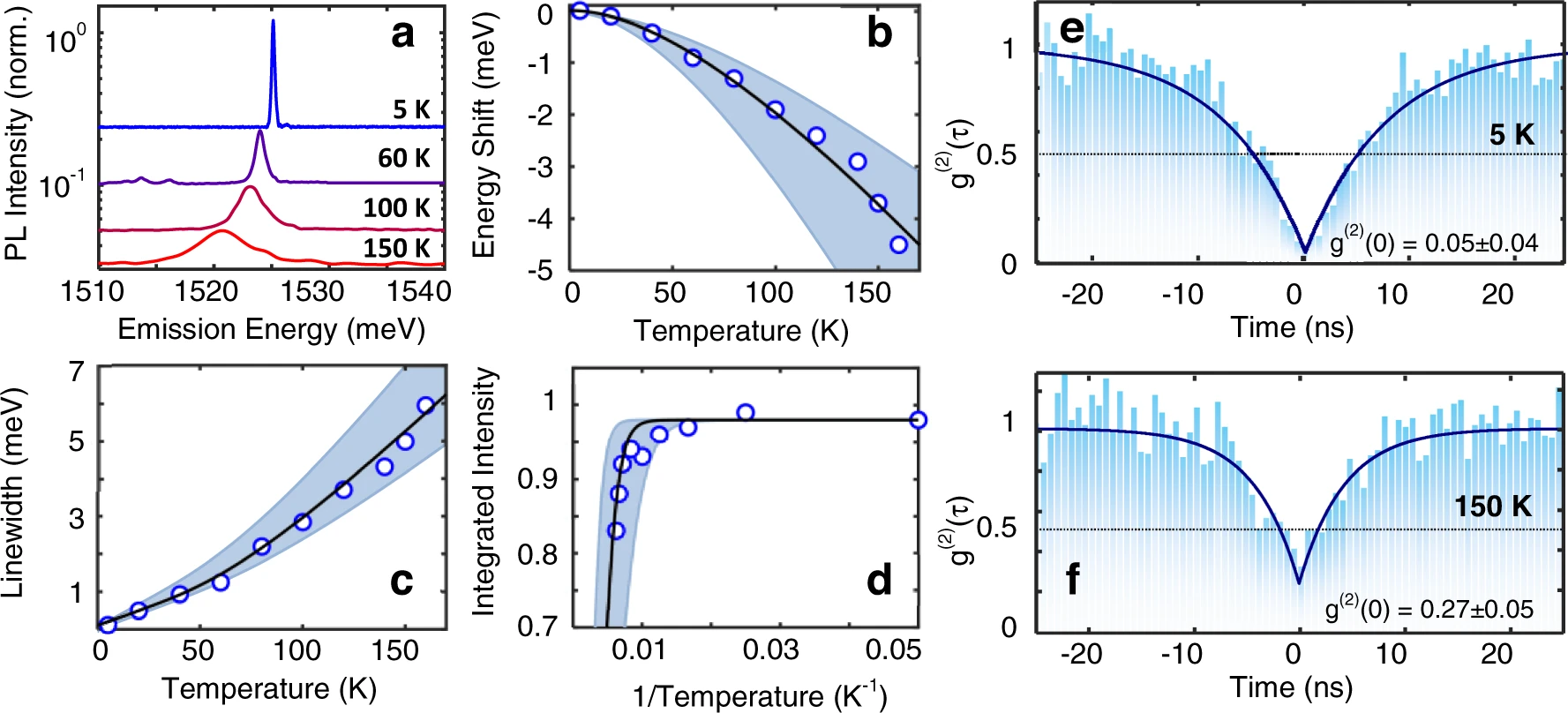}
    \caption{
        Optical emission properties of a defect in the 2D material MoSe$_{2}$ at temperatures from 5 K to 150 K \cite{Parto2021}.
    }
    \label{fig:2d-pl}
\end{figure}

Similarly to bulk color centers, the optical transitions of 2D defects are coupled to lattice phonons which cause the ZPL to broaden; see for example the temperature-resolved spectra of defects in WSe$_2$ in Fig.~\ref{fig:2d-pl}.
The phonon bath can also affect the spin degree of freedom of defects.
This can be seen in the behavior of the $T_1$ time of the V$_\text{B}^{-}$, which rapidly increases as temperature is reduced \cite{hbn_cryo_spin}.

\section{Cryocooling}
Helium has the lowest boiling point (4.2 K) among all gases, and for that reason is of central importance in cryogenics; helium is the working fluid of choice for most cryogenic systems used in quantum research.
Until the mid-20th century, cooling power for low-temperature experiments was provided in an `open-loop' manner: relatively large cryogenic plants produced liquid helium that was then brought into direct contact with the experimental apparatus.
While open-loop systems have the virtue of simplicity, they also have significant practical drawbacks.
Handling liquid cryogens requires safety precautions for researchers, and requires regular attention for monitoring and refilling.
If temperatures other than 4.2 K are needed, achieving stable and precise temperature control can be difficult with open-loop cooling.
Most importantly, the consumption of liquid helium quickly becomes enormous, and the additional equipment required to capture and re-liquefy the boil-off is frequently not available.
These problems were solved by the advent of closed-loop cryogenics: mechanical refrigeration engines using helium as the working fluid.

The original helium liquefiers were based on the Joule-Thomson effect, in which a gas is pressurized and allowed to undergo isenthalpic expansion through a narrow valve.
This produces either heating or cooling (depending on the sign of its JT coefficient).
Because gaseous helium has a negative JT coefficient above about 40 K (i.e. isenthalpic expansion heats the gas instead of cooling it), the earliest helium liquefiers required a series of different liquefied gases, each cooling the next to lower and lower temperatures, ending with cold helium gas which could then be liquefied by JT expansion.
This changed in the 1940s with the development of the Collins liquefier, which used adiabatic expansion of compressed helium against pistons to pre-cool compressed helium before expanding it in a JT valve to achieve liquefaction.
This is similar in spirit to how the enormous cryogenic plants used in particle accelerators operate even today.
While this eliminated the need for liquid cryogens other than helium, and greatly simplified access to liquid helium temperatures, the machine had numerous cryogenic moving parts, limiting its reliability.
Furthermore, the system was still open-loop: helium had to be first liquefied, then used for cooling.

In the 1960s, efforts to overcome these limitations resulted in the invention of the two main types of closed-cycle system in use today, the Gifford-McMahon (GM) cryocooler and pulse tube (PT) \cite{GiffordMcMahon}.
While their early incarnations were too inefficient and underpowered for practical use, decades of refinement led to machines which, beginning in the 1990s, could provide multiple watts of cooling power at base temperatures well below 4 K, with reliability reaching thousands of hours of operation.
In parallel, the push to achieve colder and colder temperatures culminated in the development of the dilution refrigerator (DR), also in the 1960s.
This represented a brand new approach to cooling, relying on a remarkable phase separation seen in mixtures of the isotopes helium-3 and helium-4 at extremely low temperatures.
Modern DRs are the workhorses of ultra-low temperature physics, routinely producing cooling deep into the millikelvin range.

This section details the operating principles and performance characteristics of the main cryogenic platforms in use today.
Table~\ref{tab:cryostats} summarizes their base temperatures and vibration levels for quick reference.

\subsection{Open-loop Cryocooling}
The simplest (and indeed original) method of cooling a sample is to simply immerse it in liquid helium, or to flow the helium through a pipe or chamber thermally connected to the sample.
They are known as ``wet" cryostats, since they involve direct handling of a liquid cryogen and are frequently also open-loop, meaning that the helium boiled off by heat absorbed from the sample is permitted to escape to the atmosphere.
The heat of vaporization of liquid helium-4 (which constitutes virtually all of naturally occurring helium) is 2.6 kJ/L, so a wet cryostat can achieve a cooling power of $P$ Watts at 4.2 K by consuming $1.38 P$ liters of helium per hour.

Liquid helium is typically stored in 100 L Dewars which have evaporation rates on the order of 1\% per day so, once at base temperature, experiments using open-loop flow cryostats can persist on such a Dewar for days to weeks depending on the dissipated heat load \cite{VENTURA2008xiii}.
The helium consumed in the cooldown process itself can also be very substantial.
Notably, helium's heat of vaporization is extremely low: 1/1000 that of water, and 1/60 that of liquid nitrogen.
Given the high price of helium, pre-cooling wet cryogenic systems to 77 K with liquid nitrogen before introducing helium can be advantageous \cite{PobellBook}.
It is also critical to use the cooling power of the cold helium gas during a cooldown, as the enthalpy change of helium between boiling point and liquid nitrogen temperature (77 K) is 64 kJ/L \cite{PobellBook}.
Using these techniques can reduce liquid helium consumption by orders of magnitude.
It is also possible to convert an open-loop flow cryostat into a closed-loop system by adding a helium pump and closed-cycle cryocooler to re-liquefy the gaseous helium from the exit of the cryostat.

\begin{figure}[htp]
    \centering
    \includegraphics[width=\columnwidth]{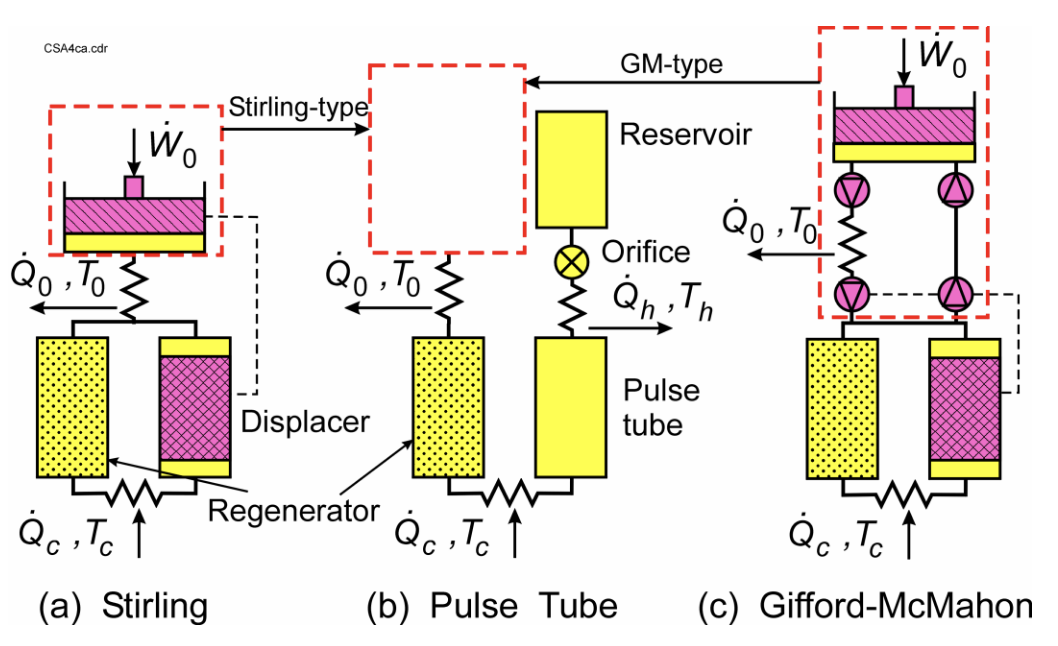}
    \caption{
        Basic schematics of Stirling, Pulse Tube, and GM closed-cycle cryocoolers.
        All three designs use oscillating pressure and gas flow to transfer heat, but differ in their mechanical implementation: the Stirling cycle uses a piston for compression, the GM cooler employs a separate displacer, and the pulse tube eliminates cold moving parts entirely.
        Adapted from \cite{Radebaugh2021}.
    }
    \label{fig:cryocooler-schematic}
\end{figure}

\subsection{4 Kelvin Closed-Loop Cryocoolers}
Fundamentally, both GM and PT cryocoolers consist of a metal tube, one end of which is connected to a helium compressor by a valve; the valve is opened and closed cyclically so that the pressure in the tube oscillates.
Within the tube is a thermal regenerator: a matrix of material with high specific heat, designed to have a large surface area so as to be in good thermal contact with the helium without creating significant resistance to its oscillating flow.
The key to cryocooler operation is that the cyclical flow of helium in and out of the tube is made to be in phase with the oscillation in its pressure.
This enables heat to flow out of the tube: the compressor end remains near room temperature, while the opposite end is cooled.
There is a close analogy with alternating-current electrical power, which only flows when current and voltage are in phase \cite{Radebaugh2021}.
The difference between GM and PT cryocoolers lies in how this phase relationship is achieved.

As illustrated in Fig.~\ref{fig:cryocooler-schematic}, both types of cryocooler bear a resemblance to the Stirling cycle, in which the pressure oscillation is produced by a piston, and a separate moving element (called a displacer) ensures the proper phase relationship between pressure and gas flow.
In GM coolers, the regenerator and displacer are unified into one element \cite{GiffordMcMahon}.
The motion of the regenerator/displacer is coupled to the opening and closing of the helium compressor valve by a mechanical linkage.
As the valve opens to the high-pressure setting, the regenerator is moved toward the warm end, displacing helium through it toward the cold end and cooling it in the process.
The valve then switches to low pressure, letting the helium undergo isothermal expansion, producing cooling power.
At the same time, the regenerator moves toward the cold end; the helium flows through it, warming as it absorbs heat from the regenerator.
Unlike the Stirling cycle, the GM cooler's pressure source is decoupled from the cycle of the cooler (typically 1-2 Hz).
This enables the use of oil-lubricated compressors operating at higher speeds and with smaller swept volumes, which are significantly more reliable \cite{Radebaugh2003}.
However, the GM machine still has at least one cryogenic moving part (the displacer/regenerator), which is usually connected to the same motor which operates the rotary helium valve via a room temperature linkage.
While the Gifford-McMahon design is quite simple and robust, the system's moving parts (valve, displacer, and linkage) induce vibrations which can be problematic for highly sensitive experiments.


Instead of a movable displacer, in a pulse tube the necessary phase relationship between gas flow and pressure is created by a small orifice inside the tube through which the helium flows as its pressure varies \cite{Radebaugh2003}.
As can be seen in Fig.~\ref{fig:cryocooler-schematic}, this allows the pulse tube to produce cryogenic refrigeration with no cold moving parts, allowing for greater reliability and longer lifetime, and eliminating most of the vibration produced by GM cryocoolers.
Many variations on the pulse tube were explored following its discovery in the 1960s, exploring the effect of changing the dimensions of the tube and its orifice, and of providing additional connections between the ends of the pulse tube.
Modern versions feature a secondary orifice which bypasses the regenerator, instead flowing to the cold end through an inertance tube whose dimensions tune the flow/pressure phase relationship, allowing for higher performance \cite{Radebaugh2003}.

The regenerator lies at the heart of both GM and PT cryocoolers, and a regenerator's ability to efficiently exchange a large amount of heat with the helium gas is a fundamental factor limiting a cryocooler's performance.
An ideal regenerator should have a large thermal mass so that it can meaningfully warm and cool the helium working gas.
Below about 20 K, as temperature $T$ falls, the heat capacity of metals like lead and copper (used in early regenerators) falls as $T^3$; meanwhile the volumetric heat capacity of the helium actually increases, since the density of the gas rises.
At some point the regenerator is overwhelmed, limiting the cryocooler's performance.
In contrast, many rare earth metals and alloys experience paramagnetic phase transitions in the 4-20 K range, causing their heat capacities to deviate from the $T^3$ Debye law.
In the 1990s, following decades of experimentation, it proved possible to fabricate reliable regenerators from rare earth compounds, and they now feature in all modern cryocoolers.
Beginning in the mid-1990s, multi-stage GM and PT cryocoolers began to reach liquid helium temperature and below \cite{Matsubara1995}, and shortly thereafter two-stage pulse tubes reaching 4 K became available from multiple vendors \cite{cryomech_pulse_tube, sumitomo_pulse_tube}.
Today closed-cycle cryocoolers with base temperatures below 2.8 K and cooling power in the 1.5 - 2 W range at 4 K are commercially available.

\begin{table}[htbp]
\centering
\label{tab:cryostats}
\small
\begin{tabularx}{\columnwidth}{l c c}
\toprule
\textbf{Type} & \textbf{Base $T$} & \textbf{Vibration}  \\ \midrule
Wet $^4$He flow    & $4.2$\,K     & Ultra-low            \\
Pulse Tube     & $\sim 3-4$\,K      & Low                \\
Gifford-McMahon & $\sim 2.5$\,K & Moderate  \\
Pumped $^4$He & $\sim 1$\,K  & Low \\
Pumped $^3$He  & $\sim 300$\,mK    & Low                  \\
Dilution refrigerator        & $\sim 10$\,mK     & Low            \\
\bottomrule\
\end{tabularx}
\vspace{2pt}
\begin{flushleft}
\caption{
    Comparison of common cryogenic refrigeration technologies and their operational characteristics.
}
\end{flushleft}
\end{table}

\subsection{1 K Closed-Loop Cryocoolers (Helium-3 cryostats)}
A common technique for achieving temperatures somewhat below 4.2 K is to first produce a pot of liquid helium and then use a vacuum pump to reduce the vapor pressure above the liquid, suppressing the boiling point.
This method was used even in Onnes' original liquefaction of helium, where he was able to reach as low as 1.5 K \cite{Onnes1909}.
The stronger the pump, the lower the temperature obtained, and the higher the fraction of helium residing in the vapor phase.
Notably, helium-3 and helium-4 have significantly different vapor pressures, so the minimum temperature achievable in this arrangement depends strongly on which isotope is used as the working fluid.
When helium-4 is used, temperatures of about 1 K can be achieved; below this point its vapor pressure (and thus available cooling power) drops extremely rapidly.
At a given temperature, the vapor pressure of helium-3 is much higher (about 14 times higher at 1 K), and cryostats pumping on helium-3 vapor can reach as low as 0.3 K.
Until the development of dilution refrigerators, this type of cryostat was the dominant method of reaching the sub-Kelvin regime.
Although they are now somewhat less common than 4 K direct-cooling cryostats or dilution refrigerators, a number of vendors still manufacture them.

To produce the pot of liquid helium, these cryostats typically use Gifford-McMahon cryocoolers due to their high cooling powers; a Roots or scroll pump is then used to reduce the vapor pressure.
The low temperatures offered by these cryostats make them well-suited to hosting quantum devices that incorporate SNSPDs, which typically have transition temperatures around 2.5 K or below \cite{Norman2025}.
This type of cryostat also typically provide large cooling powers, which can be beneficial for large, integrated quantum systems, or those that dissipate heat.
This heat can come, for example, from bright pump light, active photonic switches, or the use of electrical current to drive electroluminescence from color centers or 2D materials.

\begin{figure}[htp]
    \centering
    \includegraphics[width=\columnwidth]{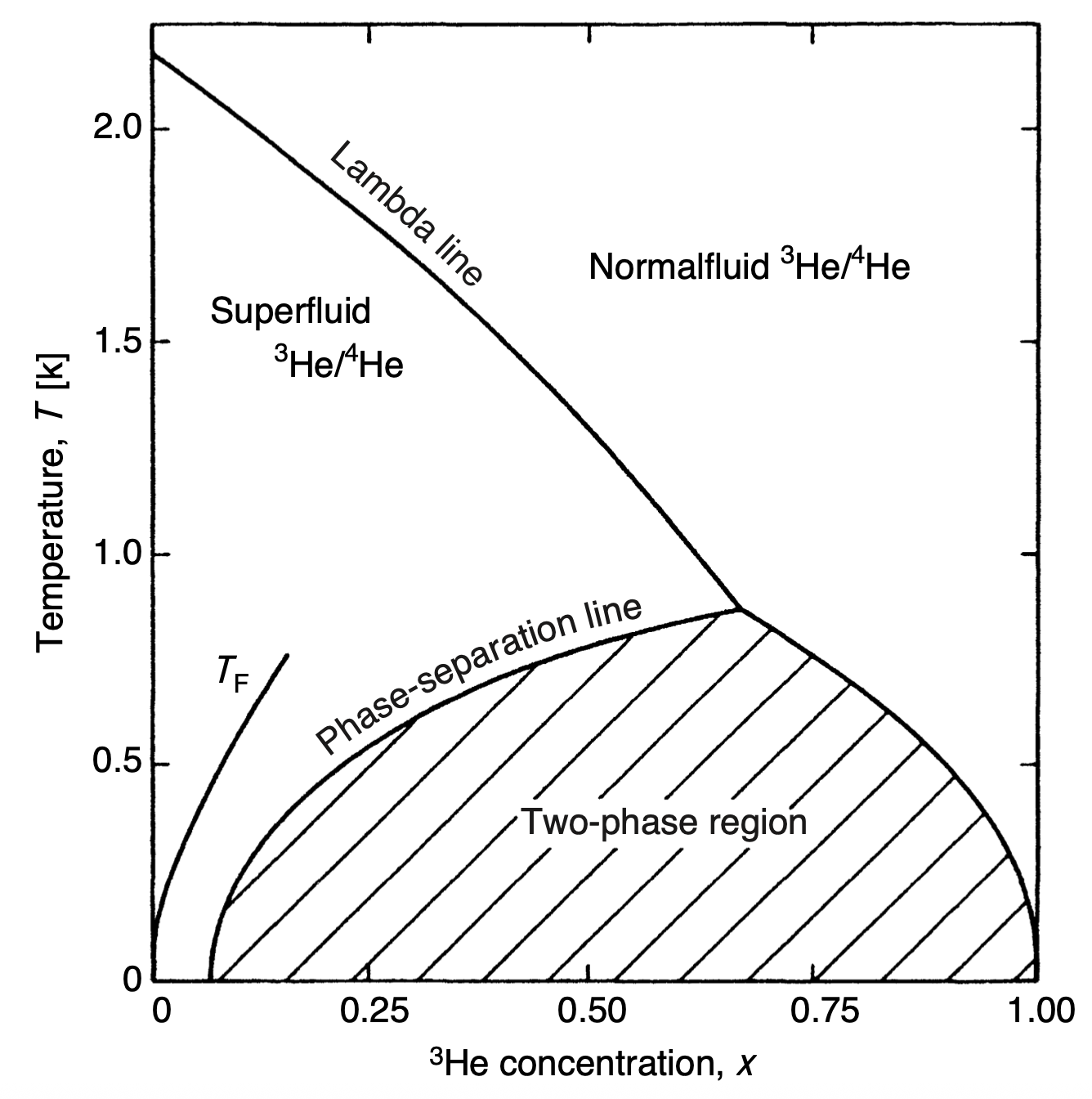}
    \caption{
        Phase behavior of $^3\text{He}\text{-}^4\text{He}$ liquid mixtures, showing the transition from a homogeneous normal fluid to a phase-separated superfluid state.
        The lambda line represents the second-order superfluid transition, which meets the first-order phase-separation curve at the tricritical point ($T \approx 0.87\text{ K}$, $x \approx 0.67$).
        Below the phase-separation boundary, the mixture spontaneously segregates into a $^3\text{He}$-rich phase (concentrated) and a $^4\text{He}$-rich phase (dilute).
        Notably, the dilute phase retains a finite $^3\text{He}$ solubility of approximately $6.6\%$ even as $T \to 0\text{ K}$, a property fundamental to the operation of dilution refrigerators.
        Reproduced from \cite{Pobell2007_Ch7}.
    }
    \label{fig:helium_phase}
\end{figure}

\subsection{Millikelvin cryostats (dilution refrigerators)}
Reaching the deep sub-Kelvin regime is challenging, partly because most sources of cooling power have already been exhausted at these low temperatures.
Dilution refrigerators (often called `dil fridges' for short) achieve this by exploiting the existence of a phase transition in mixtures of helium isotopes.
As shown in Fig.~\ref{fig:helium_phase} at very low temperatures a mixture of helium-3 and helium-4 has two phases: one which is rich in helium-3 and another which is dilute (and in which the helium-4 is superfluid).
Below about 0.87 K, the two phases spontaneously separate, with the less-dense rich phase floating on top of the dilute phase.
Movement of helium-3 atoms across the phase boundary (from rich to dilute) is an endothermic process, driven by heat absorbed from the environment.
In a loose analogy, helium-3 in the rich phase can be thought of as ``evaporating" into the superfluid helium-4 of the dilute phase below, cooling the cryostat as it does so.
In order to produce continuous cooling, the rich phase at the top of the mixing chamber must be continually replenished, and the helium-3 in the dilute phase must be removed.
The dilute phase at the bottom of the mixing chamber is in communication with a still which is positioned in a higher-temperature stage of the DR.
Because helium-3 has a much higher vapor pressure than helium-4, the gas above the still is mostly helium-3.
A vacuum pump collects this gas, which is recondensed and circulated back to the rich end of the mixing chamber to close the loop.
As the temperature falls toward absolute zero, the fraction of helium-3 in the rich phase tends toward 100\% while, importantly, the solubility of helium-3 in the dilute phase does not tend toward zero (it levels off at $\sim 6.5\%$).
This fortuitous fact allows the dilution refrigerator to, in principle, continue providing cooling power arbitrarily close to absolute zero.
In practice, DRs typically have no-load base temperatures of around 10 mK and are by far the most common type of refrigerator for work in this regime.

\begin{figure}[htp]
    \centering
    \includegraphics[width=\columnwidth]{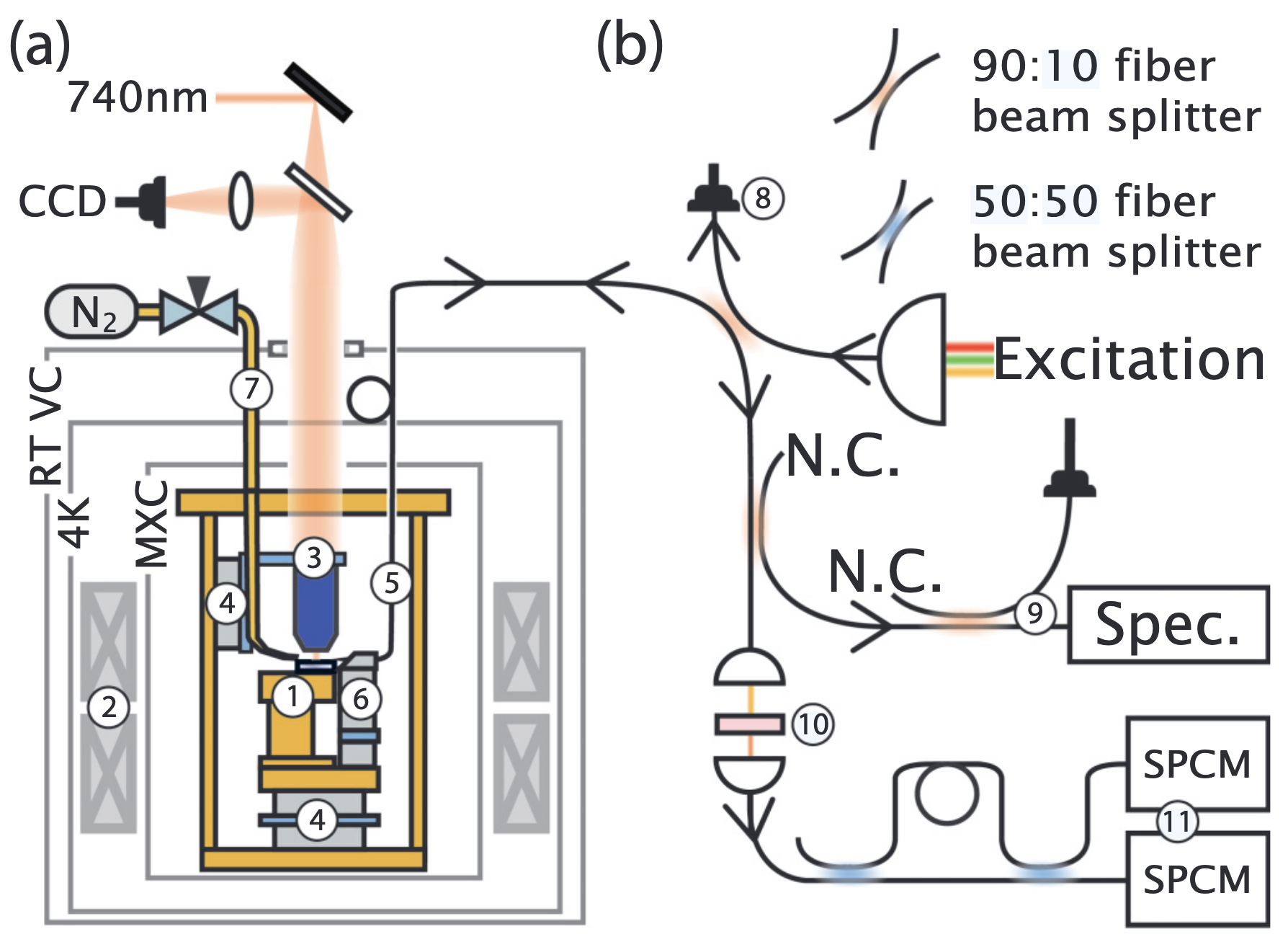}
    \caption{
        Diagram of an experiment on SiV centers in diamond. The sample rests on a piezo nanopositioner inside a DR; a second nanopositioner holds a tapered fiber which is used to couple to the waveguides on the sample.
        A wide-field imaging system uses a beam path vertically down through the top of the cryostat, where a cryogenic microscope objective is focused on to the sample.
        Note the superconducting magnet mounted in the 4 K part of the cryostat.
        Reproduced from \cite{PhysRevB.100.165428}.
    }
    \label{fig:optical-dil-fridge}
\end{figure}

The cold space of a dilution refrigerator consists of a series of several stages descending from room temperature down to the millikelvin range, typically with the warmest at the top and coldest at the bottom.
To liquefy the helium-3/helium-4 mixture so that dilution can begin, a primary source of cooling is required.
From its invention in the 1960s, the original DRs were ``wet": they were pre-cooled directly with a supply of liquid helium, often in an open-loop configuration \cite{PhysRev.128.1992, HALL196680}.
With the advent of powerful sub-4 K cryocoolers in the 1990s DRs went ``dry", eliminating liquid cryogen pre-cooling in favor of closed-cycle cryocoolers \cite{UHLIG1993}.
Pulse tubes are preferred over GM coolers for this purpose, because the higher vibrations from GM devices shake the mixing chamber, disrupting the dilution process and limiting performance \cite{UHLIG2002}.
In modern DRs, the use of a pulse tube for precooling is the de facto standard.
Pulse tubes generally require two stages to get cold enough to condense the helium in the mixing chamber; the high cooling powers available especially at the first stage of the PT (usually at around 50 K) provide a convenient way to thermally anchor wires and fibers entering the cryostat from the room temperature side, helping to thermally isolate the millikelvin sample space.
The cooldown time for a DR depends on the size of the apparatus to be cooled and the available cooling power; today DRs are commercially available with sample spaces ranging from centimeters to half a meter in diameter; Fig.~\ref{fig:optical-dil-fridge} shows an example schematic of how a dilution refrigerator can be configured for quantum optics experiments.
Larger machines sometimes feature multiple pulse tubes and mixing chambers, and can take as long as 24 hours to reach base temperature.
In some systems it is possible to flow liquid nitrogen through the upper stages to jump-start the cooling before allowing the pulse tube to take over, which can shorten the overall cooldown time.

\section{Cryostats for Optical Quantum Science}

The needs of optical quantum research place specific constraints on the design of cryostats.
The temperature required by the sample (and the heat load it generates) largely determines the type of refrigeration, and the type of access to the sample needed constrains the design of the cryostat's shielding.

Optical access to the sample space has always been an important design criterion in cryogenic systems; even Onnes' first liquefaction of helium in 1908 was confirmed by visually observing the liquid in the glass vial in which it was condensed.
Much like in Onnes' time, modern flow cryostats still benefit from high cooling powers and can offer optical access simply by placing the sample close to a window in the vacuum housing.
Since the early 2000s, fully integrated closed-cycle cryostats reaching 4 K and below have become commercially available.
These systems' low maintenance needs and lack of liquid cryogens have made them very popular.
Their ability to operate stably for long durations with precise temperature control and large sample spaces has made it easy to integrate bulk optics and sample mounts into the cryogenic space, enabling new techniques for optical coupling to samples.

A typical cryostat design has a thick metal vacuum shroud as its outermost layer, followed by one or more thinner thermal radiation shields nested within.
These shields surround the sample space in a concentric arrangement, with temperatures progressively decreasing from room temperature at the vacuum shroud to the cryostat's base temperature at the center.
Typically constructed from highly polished aluminum or copper for their excellent thermal conductivity and low emissivity, these radiation shields are often thermally linked to intermediate stages of the cryocooler.
This configuration leverages excess cooling capacity available at temperatures above the system's base temperature.

\subsection{Optical access}
Free-space optical access can be enabled by providing for a window in each layer of thermal shielding.
Importantly, these windows cannot be simple holes but must incorporate a physical material (commonly sapphire or fused silica), so that thermal radiation does not pass through.
Anti-reflective window coatings are also often important when dealing with single-photon emitters, which can be quite dim.
Providing unobstructed beam paths between room temperature and the sample space can be challenging for some cryostat designs, particularly dilution refrigerators and cryostats which incorporate large magnet coils.
In these setups, vertical access is typical: room-temperature optics are often placed at the top of the cryostat, and the beam is finally directed downward into the sample chamber.
A vertical beam path of this type is illustrated in Fig.~\ref{fig:optical-dil-fridge}, where a wide-field imaging system enters through the top of the dilution refrigerator.
While this arrangement allows retaining the relative simplicity of free-space access, alignment as a practical matter can be somewhat more difficult.

\begin{figure}[htp]
    \centering
    \includegraphics[width=\columnwidth]{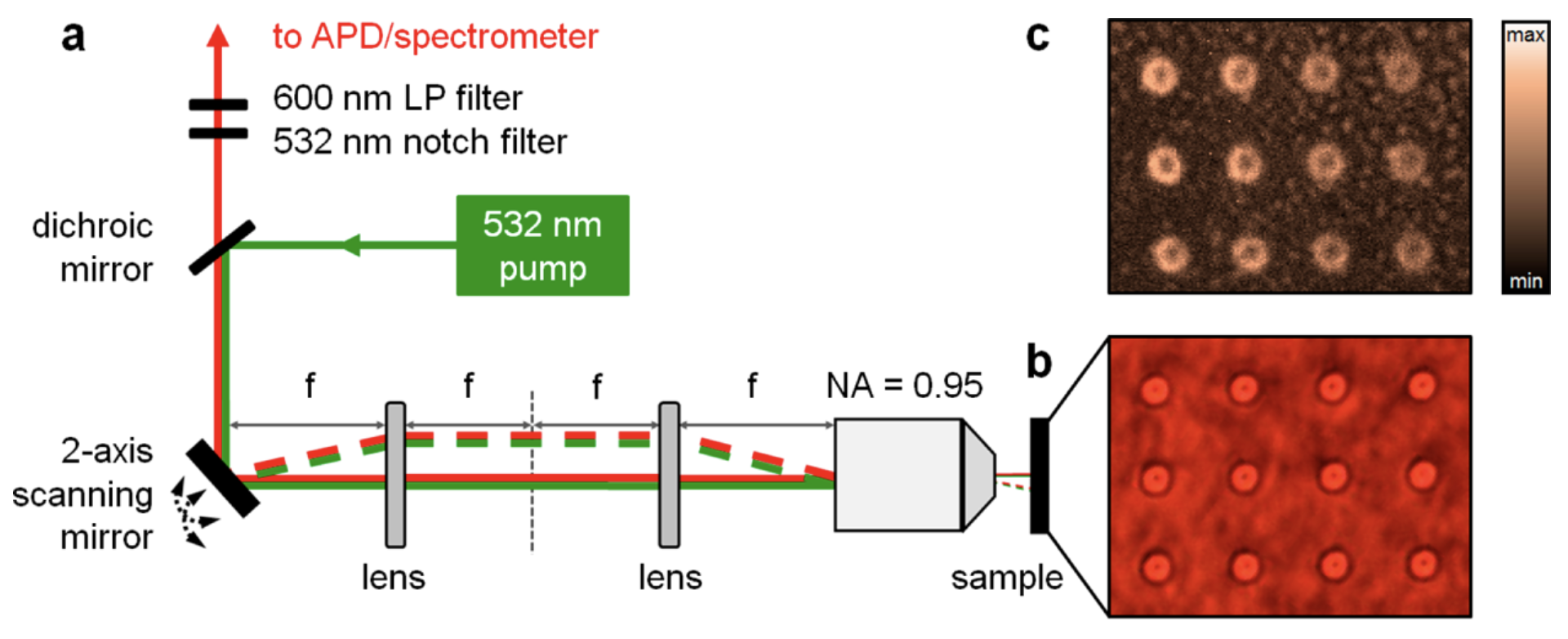}
    \caption{
        (a) Diagram of a typical scanning confocal laser microscope setup used to study quantum emitters and nanophotonic devices; (b) optical microscope image of an array of silicon carbide microdisk resonators (c) 2D photoluminescence scan taken with the confocal laser microscope.
        Adapted from \cite{Radulaski2015}.
    }
    \label{fig:4f-setup}
\end{figure}

When coupling to a sample by free-space, microscope objectives are generally used to focus on the sample.
For 4 K or 1 K cryostats (which generally require fewer layers of thermal shielding than dilution fridges) this can be done by simply positioning the sample close to a window inside the cryostat, and using a room-temperature microscope objective with a long working distance on the other side.
However, this working distance requirement can place limitations on the NA and magnification which can be achieved.
Another drawback to this arrangement is that proximity to the thermal shielding can mean the sample is not at the base temperature of the cryostat, which can be a concern if very precise temperature control is required.
Objectives designed for compatibility with cryogenic temperatures are also commercially available.
While often more expensive, placing a cryogenic objective inside the sample chamber can side-step the drawbacks of a room-temperature objective \cite{10.1063, Sapienza2015}.

Quantum emitters are often fabricated by implantation or growth processes, and therefore have a random spatial distribution; it can be necessary to explore large regions of a sample in search of functional emitters.
One solution to this is to use piezoelectric nanopositioners, which can provide as much as a centimeter of displacement with nanometer-precision at deep cryogenic temperatures.
The simplest arrangement is to mount a sample on one or more nanopositioners, allowing it to be translated in up to three dimensions, as well as rotated.
Modern cryostats often feature spacious chambers that can accommodate multiple piezoelectric nanopositioners, allowing for more complex setups where multiple electrical or fiber-optic probes have access to wide areas of the sample.
However, it is important to note that nanopositioners can shift during cooldown.
In setups where the sample is close to a window or objective, it is prudent to move it a safe distance away to avoid collision.

\begin{figure}[htp]
    \centering
    \includegraphics[width=\columnwidth]{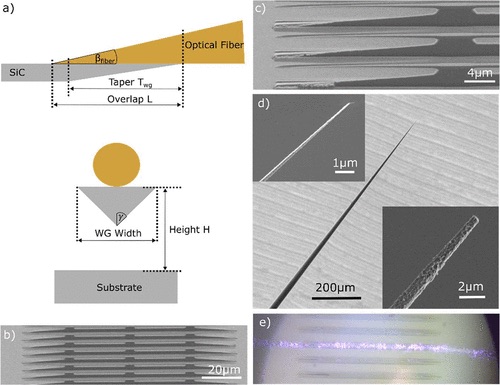}
    \caption{
        A waveguide-fiber interface made from tapered silicon carbide.
        Tapered interfaces like this provide low-loss coupling between on-chip waveguides and fiber optics, enabling efficient extraction of single photons from integrated quantum devices.
        Reproduced from \cite{Krumrein2024}.
    }
    \label{fig:fiber-interface}
\end{figure}

In more space-constrained setups, a 4f confocal microscope arrangement can allow for scanning a free-space probe beam across the surface of the sample without any cryogenic movement.
A scanning confocal microscope can be formed by placing a galvo mirror, a pair of lenses, and a pinhole aligned with a microscope objective.
As the galvo mirror tilts the angle of the beam, the lenses and objective convert that tilt into a position on the sample.
This makes it easy to probe different regions without needing to physically move the sample inside the cryostat, and is commonly used to make 2D maps of the intensity of emission across the sample \cite{Norman2025, diamond_nv_confocal, acsphotonics.0c00218}.
Figure~\ref{fig:4f-setup} shows an example scanning confocal microscope setup and the 2D photoluminescence maps it can produce.

When working with quantum emitters in nanophotonic structures such as waveguides, light is required to travel in the plane of the sample, which conflicts with conventional out-of-plane free-space optical access.
A simple solution to this is edge coupling, in which an edge of the sample is highly polished and light is directed into the device from its facet (typically with a lensed fiber tip), enabling efficient in-plane light insertion \cite{app10041538, Wan2020}.
An alternative approach that can achieve higher coupling efficiency is the use of tapered waveguide-fiber interfaces, where a tapered section of the on-chip waveguide is brought into close proximity with a tapered optical fiber (see Fig.~\ref{fig:fiber-interface}), allowing evanescent coupling between the two modes.
Using fiber as the optical interface can also be beneficial in setups where obtaining a clear free-space optical path is challenging, although this usually necessitates a larger sample chamber to allow room for nanopositioners holding the fiber. 
Figure~\ref{fig:optical-dil-fridge} demonstrates such an arrangement, with piezoelectric nanopositioners positioning both the sample and a tapered fiber for waveguide coupling.
Vacuum-tight pass-throughs are used to run fiber into a cryostat from the room-temperature side; the intrinsically low thermal conductivity of glass fiber allows for low parasitic heat load on the sample space.

\begin{figure}[htp]
    \centering
    \includegraphics[width=\columnwidth]{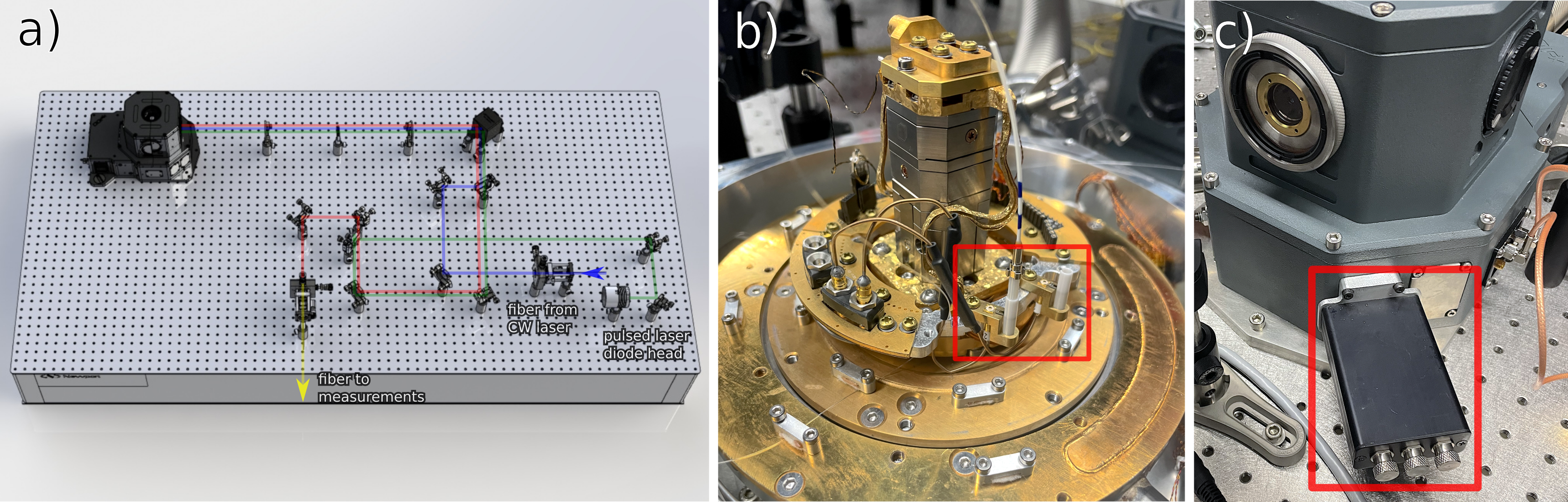}
    \caption{
        Example of integrating bulk SNSPDs into a 1.6 K cryostat. (a) Beam path for a 4f confocal optical setup interfaced with the cryostat. (b) View inside the cryostat sample chamber showing nanopositioners and SNSPDs (red box). (c) Vacuum-tight fiber ports cryostat, providing fiber-coupled optical interface to the SNSPDs. Adapted from \cite{Norman2025}.
    }
    \label{fig:icecap_snspds}
\end{figure}

When base temperatures are low enough (below 2 K) and available cooling power is high, it is possible to integrate SNSPDs directly into the sample chamber \cite{Norman2025}.
An example of this integration is shown in Fig.~\ref{fig:icecap_snspds}, where bulk SNSPDs are mounted in the sample area of a 1.6 K cryostat.
This can be an efficient arrangement, eliminating the need for a separate cryostat for the detectors.
Furthermore, when testing samples which interface to a fiber, the fiber can route directly to the SNSPDs without exiting the vacuum shroud.
This minimizes the length of fiber exposed to room-temperature blackbody radiation, effectively lowering the background count rate of the detectors.

Beyond placing packaged detectors in the sample space, SNSPDs can also be fabricated directly onto a nanophotonic chip.
These on-chip detectors are typically formed by depositing the superconducting wire on top of a waveguide; photons within the waveguide evanescently couple to the detector.
Designs like this can achieve near-unity internal detection efficiency while bypassing the insertion losses inherent to fiber-to-chip coupling \cite{psiq_manufact}.
While chip-integrated detectors offer a path to highly scalable quantum circuits, they also introduce cryogenic design considerations regarding heat flow on the chip.
Coax cables for the detectors must be thermally lagged to prevent heat leakage from the environment, and heat generated elsewhere on the chip (from laser pump pulses, for example) must be dissipated to avoid warming up the SNSPDs.

It is also possible to redirect out-of-plane light into the plane of the sample (and vice versa) with grating couplers, which are diffractive nanophotonic structures \cite{saha2025triangular, mi11070666, 7725557}.
This allows for probe or excitation beams to be redirected laterally to interact with emitters in a sample and conversely, for quantum light produced in on-chip waveguides to be directed upward toward collection optics.
The free-space end of a grating coupler can interface with a microscope objective, or a fiber tip positioned directly above the grating coupler.
When interfacing with a fiber, piezoelectric nanopositioners are typically used to give fine control over the alignment of the fiber relative to the sample.


\section{Trends and Future Directions}
\subsection{Scalability}
Scaling up quantum technologies to sizes useful for practical applications continues to be a longstanding challenge.
For example, it has been estimated that cryptographically relevant factorization using Shor's algorithm may require about one million physical qubits \cite{gidney2025}.
This stands in stark contrast to current quantum processors which contain about a thousand qubits at most (for superconducting systems) \cite{CarreraVazquez2024, castelvecchi2023ibm} and many fewer for optical systems \cite{Taminiau2014, PhysRevB.100.165428, Parker2024}.
For quantum computers which require cryocooling, physical space inside the cryostat represents one limitation to scaling.
This constraint is most acute for dilution refrigerators, which have significantly larger physical plants and power requirements per unit of sample space compared to 4 K or 1 K systems.
Several groups have already constructed large milliKelvin cryostats by combining as many as 10 dilution units in parallel providing as much as 500 $\mu$W of cooling power in the milliKelvin range \cite{Gumann_Chow_2022, Bluefors_KIDE, Hollister_2022}.

At very large scale, cryostats can benefit from abandoning direct mechanical cryocooling in favor of central cryogenic plants reminiscent of those used to cool the superconducting magnets in particle accelerators, such as the LHC at CERN.
These systems work by pumping large volumes of cold, supercritical helium to the point of use, where it is then cooled further by expansion.
This approach has been used in accelerators to cool enormous apparatus to temperatures below 2 K \cite{cryogenicsAtCern}, which is compatible with many solid-state quantum systems \cite{psiq_manufact, SimmonsPhotonicArchitecture}.
For large DRs, this supply of liquid helium can be used in the pre-cooling stages above the mixing chambers \cite{Hollister_2022}.


\subsection{Cryogenic probe stations}
A key step in production of classical semiconductor products is automated testing.
Very advanced systems have been developed which can load an undiced wafer from a cassette and rapidly conduct in depth electrical, optical, and visual inspection of each of the thousands of components on the wafer.
This validation allows not only quality control of the huge mass of electronics produced, but also provides feedback to the earlier steps in the fab process.
As quantum systems grow in scale, test systems must grow in tandem.
However, the cryogenic temperatures required by some quantum technologies creates additional hurdles to automated testing: wafers are quite large compared to typical cryostat sample spaces, and minimizing the time to cool incoming wafers from room temperature is challenging.


In the 1980s, motivated by the desire to test superconducting electronics, IBM demonstrated a semi-automated probe station for two-inch wafers based on a dipstick cryostat \cite{prober1983}.
A wafer would be loaded onto the chuck at room temperature and the probe card manually aligned to the pattern of test pads.
The probe station would then be lowered into a Dewar of liquid helium, which was wrapped in mu-metal shielding to dampen magnetic noise.
Stepper motors at room temperature then moved the probe card from device to device.
Position accuracy was estimated to be roughly 50 $\mu$m.
A similar cryostat was also built by the Fujitsu company in 1986 relying on fully manual control of the probe card, which necessitated a fiber-based imaging system to guide the user.

An automated 1 K probe station for 300 mm wafers was recently demonstrated, 
which has been used to characterize silicon spin qubits \cite{blueforsProber}.
The system can cool a wafer from room temperature to 1 K in 2 hours, with total turnaround time between measurements of 4-5 hours.
This capability has enabled high-volume statistical characterization of silicon spin qubit devices across full 300 mm wafers, demonstrating the role of cryogenic wafer-scale probing in optimizing CMOS-compatible fabrication processes for quantum technology \cite{Neyens2024}.
Since the sample space is quite large, dedicated pulse tubes cool the radiation shields, and a separate helium-vapor cryostat is used to cool the wafer chuck.
A load-lock system allows wafers to enter and leave the probe station while allowing the main chamber to remain at cryogenic temperatures and under vacuum, which is critical to avoid the long cycle times for such a large cryostat.

A 4 K wafer prober (for 150 and 200 mm wafers) was also recently demonstrated \cite{hpdProber}.
This device includes both passive and active suppression of magnetic field noise by multiple layers of mu-metal shielding in addition to a magnetic coil.
This reduces the field inside the probe station to 100 nT.
Interestingly, the cryogenic system is open-loop, relying on two liquid helium Dewars (although the boil-off can in principle be recondensed by use of a cryocooler).
As with many flow cryostats, the cooling power of this system is very high, and it can bring a wafer from room temperature to 4 K in just 10 minutes.
This system has been used to characterize superconducting integrated circuits at the wafer scale, including Josephson junctions and SQUID-based magnetic field sensors \cite{westTolpygo2022wafer}.

\subsection{Energy efficiency}
Given the large amount of infrastructure required by quantum technologies, a good point of comparison with classical computing can be drawn from classical data centers.
In a typical data center, roughly half the energy consumption is due to the cooling systems, while only about a third of the energy is used by the actual computers and networking equipment \cite{ALKRUSH2024246}.

This disparity can be expected to increase for large quantum systems.
Owing to the much lower operating temperature of cryogenic cooling, the maximum achievable COP is vastly lower than the cooling used for classical systems.
In addition, quantum information processing operations themselves dissipate negligible energy, and the energy consumed by the classical control electronics is also relatively small.
As a consequence cryogenic cooling will almost certainly be the dominant use of power in a large deployed quantum computer \cite{9827605}.



\subsection{Helium-3 shortage}
For quantum technologies requiring deep sub-Kelvin temperatures, dilution refrigerators are by far the most mature cryogenic platform.
However, their dependence on the rare isotope helium-3 is potentially a major issue, as the supply of this isotope is deeply intertwined with international nuclear security concerns.
Helium has two stable isotopes: $^3$He and $^4$He; naturally occurring helium is almost entirely $^4$He, with about 2 ppm of $^3$He.
While this tiny fraction of $^3$He can be separated out from natural helium (itself mostly found co-occurring with natural gas deposits), the source of most $^3$He is aging nuclear weapons which contain tritium ($^3$H).
Tritium converts by beta decay into $^3$He, which accumulates as an unwanted waste product and must be pumped out periodically during maintenance.
Since the early 1990s global tritium stockpiles have declined, leading to a reduction in $^3$He availability \cite{science_helium}.
Nuclear security also drives the demand for scarce helium-3 due to its use in fast-neutron detectors, which are deployed to screen for the presence of radioactive material, and have been installed extensively since the year 2000 \cite{shea2010helium}.
This combination of reduced supply and enhanced demand has driven up helium-3 prices in recent years and squeezed the cryogenics market.

The $^3$He shortage is forecast to reach 55,000 L per year in the near future \cite{Verkhovniy2022}.
For comparison, a typical dilution refrigerator needs about 18 L of helium-3, though this is recycled indefinitely in the absence of leaks.
Some government agencies, such as the United States Department of Energy, have authorized production of new tritium, which in turn has bolstered supply of $^3$He.
There are even proposals for mining helium-3 from the surface of the moon, which contains a relatively high abundance of the isotope \cite{FA200715}.

\section{Summary and Outlook}

The evolution of cryogenic engineering from manual liquid cryogen handling to automated closed-cycle systems has fundamentally transformed quantum research.
As detailed throughout this review, a diverse array of platforms ranging from 4 K pulse-tube refrigerators to sub-10 mK dilution refrigerators now provide the thermal environments necessary for solid-state quantum emitters and superconducting detectors to achieve their promised revolutionary performance.
The choice of cryogenic platform depends critically on the specific quantum system: while established candidates like diamond NV centers and InAs quantum dots perform admirably at liquid helium temperatures ($\sim$4 K), emerging platforms such as group-IV defects in diamond and bound-exciton centers in silicon benefit from the more stringent sub-Kelvin environments that dilution refrigerators can best provide.

The future development of quantum technologies will increasingly be defined by scalability beyond laboratory demonstrations.
Current estimates suggest that cryptographically relevant applications may require upward of one million physical qubits, several orders of magnitude more than are currently assembled in any one system.
This massive expansion presents formidable challenges across multiple fronts.
The physical infrastructure needed to house and cool such systems will dwarf current capabilities, potentially necessitating the use of optical networking to coherently connect separate cryogenic systems, or even a transition away from individual cryostats towards centralized cryogenic plants reminiscent of those serving particle accelerators.
Parallel to this expansion in scale, the manufacturing pipeline for quantum devices must mature substantially.
Automated cryogenic wafer-scale probe stations—already demonstrated for 300 mm silicon wafers—will become essential for rapid characterization and quality control during fabrication.
These systems enable testing of thousands of devices per wafer while providing critical feedback to upstream fabrication processes, mirroring the quality assurance infrastructure that underpins classical semiconductor manufacturing.

Deployment of quantum technologies also faces significant cryogenics-related challenges.
Energy efficiency represents a critical consideration: the thermodynamic penalty for operating at cryogenic temperatures is severe, with even modest cooling powers requiring kilowatts of input power at room temperature.
As quantum data centers emerge, minimizing this overhead through optimized cryogenic architecture and careful thermal management will be essential for sustainable deployment.
A second challenge stems from the helium-3 supply chain.
With global availability tied closely to nuclear security programs, scarcity of this isotope could constrain widespread deployment of dilution refrigerators.
While various proposals for additional supply have been advanced, helium-3 availability poses a persistent challenge for technologies requiring deep sub-Kelvin operation.

Despite these challenges, the trajectory is encouraging.
The maturation of closed-cycle systems over the past three decades demonstrates that technical barriers can be overcome, while ongoing innovations in regenerator materials, vibration isolation, and optical integration continue to expand cryogenic capabilities.
For the practicing quantum scientist, understanding modern cryogenic systems is essential for informed platform selection.
The trade-offs between base temperature, optical access, magnetic field compatibility, and operational complexity must be carefully weighed.
By providing this technical foundation, we hope this review empowers researchers to navigate the cryogenic landscape and integrate these systems into next-generation quantum photonic technologies.

\section*{Acknowledgments}

We acknowledge support from the National Science Foundation CAREER award (No. 2047564).
This work was supported in part by the UC Davis 2025-26 Summer Graduate Student Research Award.

\bibliography{sn-bibliography}

\end{document}